\documentclass[a4paper,twocolumn,11pt,accepted=2020-03-24]{quantumarticle}
\pdfoutput=1
\usepackage{amsmath}
\usepackage{amssymb}
\usepackage{graphicx}
\usepackage{hyperref}
\usepackage[T1]{fontenc}
\usepackage[english]{babel}
\usepackage{placeins}
\usepackage[utf8]{inputenc}
\usepackage{mathtools}
\usepackage{verbatim}
\usepackage{tikz}
\usepackage[numbers,sort&compress]{natbib}

\DeclareMathOperator{\Tr}{Tr}

\newcommand{\braket}[2]{\langle #1|#2\rangle}

\newcommand{\ket}[1]{|#1\rangle}
\newcommand{\bra}[1]{\langle#1|}
\DeclareMathOperator{\rhont}{\rho_t^{(n)}}

\newcommand{\be}{\begin{equation}}
\newcommand{\ee}{\end{equation}}
\newcommand{\bea}{\begin{eqnarray}}
\newcommand{\eea}{\end{eqnarray}}

\begin{document}

\title{Logarithmic growth of local entropy and total correlations in many-body localized dynamics}

\author{Fabio Anza}
\email{fanza@ucdavis.edu}
\orcid{https://orcid.org/0000-0003-0695-8220}
\affiliation{Complexity Sciences Center,
Physics Department,
University of California at Davis,
One Shields Avenue, Davis, CA 95616}

\author{Francesca Pietracaprina}
\orcid{https://orcid.org/0000-0002-8927-3923}
\email{pietracaprina@irsamc.ups-tlse.fr}
\affiliation{Laboratoire de Physique Th\'eorique, IRSAMC, Universit\'e de Toulouse, CNRS, UPS, France}

\author{John Goold} 
\orcid{https://orcid.org/0000-0001-6702-1736}
\email{gooldj@tcd.ie}
\affiliation{School of Physics, Trinity College Dublin, College Green, Dublin 2, Ireland.}

\begin{abstract}
The characterizing feature of a many-body localized phase is the existence of an extensive set of quasi-local conserved quantities with an exponentially localized support. This structure endows the system with the signature logarithmic in time entanglement growth between spatial partitions. This feature differentiates the phase from Anderson localization, in a non-interacting model. Experimentally measuring the entanglement between large partitions of an interacting many-body system requires highly non-local measurements which are currently beyond the reach of experimental technology. In this work we demonstrate that the defining structure of many-body localization can be detected by the dynamics of a simple quantity from quantum information known as the total correlations which is connected to the local entropies. Central to our finding is the necessity to propagate specific initial states, drawn from the Hamiltonian unbiased basis (HUB). The dynamics of the local entropies and total correlations requires only local measurements in space and therefore is potentially experimentally accessible in a range of platforms.
\end{abstract}

\maketitle

The study of transport properties of quantum systems is a topic of paramount importance in condensed matter physics. A crucial aspect is the presence of disorder due to defects and irregularities in the material under study. In a celebrated work~\cite{Anderson1958} Anderson showed how the presence of strong disorder can completely suppress transport of non-interacting electrons in a tight-binding model. Understanding the fate of this localisation phenomenon in the presence of interactions has seen an unprecedented revival in recent years~\cite{alet2018many,bloch_review}.  In a seminal contribution Basko et. al.~\cite{Basko2006} argued that such phenomenon is stable when interactions between particles are introduced, showing the existence of a new dynamical phase of matter, the \emph{Many-Body Localized} (MBL) phase~\cite{Nandkishore2015,altman2015universal,altman2018} which, like its single particle counterpart exhibits a lack of both transport and thermalization~\cite{altman2018}. From the experimental perspective, signatures of MBL physics have recently been observed in a number of different laboratories in cold atoms~\cite{exp_1,exp_2,exp_3}, ion traps~\cite{exp_4} and NMR~\cite{exp_5}.

 As the system fails to thermalize, local observables retain memory of their initial conditions. In the last ten years there has been a large amount of effort devoted to the understanding of the MBL phase~\cite{Vosk2015, Abanin2017, alet2018many, Scardicchio2017, Nandkishore2015, Altman2015, Imbrie2017}. The defining feature of an MBL state (e.g. as opposed to Anderson localization) has been identified in the fact that, while the transport of energy and local quantities is suppressed~\cite{knapdemler,Znidaric2016,Varma2017a,Luitz2016a,luitz2017ergodic,vznidarivc2017dephasing,mendoza2018asymmetry,schulz2018energy}, there is transport of quantum information, occurring on a logarithmic time scale manifested in the growth of the half-chain entanglement entropy~\cite{Znidaric2008,Bardarson2012,Serbyn2013}. This behaviour can be explained by the emergence of an extensive set of \emph{quasi-local} integrals of motions (Q-LIOMs)~\cite{Serbyn2013,huse2014phenomenology,Ros2015,Imbrie2017}. Such objects have a support which is exponentially localized, the localization length being $\xi$. In the high-disorder regime the tails become more suppressed and the Q-LIOMs approach local quantities.

The canonical models to study MBL, which we will also use in this letter, is the XXZ spin chain with random fields:
\begin{equation}
H = \sum_{i=1}^L \left( s^x_i s^x_{i+1} + s^y_i s^y_{i+1} +  \Delta s^z_i s^z_{i+1}\right) + \sum_{i=1}^L h_i s^z_i \,\,\, , \label{eq:XXZ}
\end{equation}
where $s^\alpha=1/2\,\sigma^\alpha, \, \alpha=x,y,z$ are $1/2$-spins, the fields $h_i$ are random variables with uniform probability distribution in $\left[ -W,W\right]$ and $W \in [0,\infty)$ is the disorder strength.
For zero disorder this model is the quantum Heisenberg model; it has a many-body localization transition at $W_c\approx3.72$~\cite{Luitz2016a} at $\Delta=1$ and conserves the total magnetization $s^z=\sum_i s^z_i$.
The emergence of integrability corresponds to recasting \eqref{eq:XXZ} into the effective Hamiltonian
\begin{equation}
H_{\mathrm{MBL}}=\sum_{j_1} \lambda_{j_1}^{(1)}\tau_{j_1}^z + \sum_{i_1<i_2} \lambda_{i_1 i_2}^{(2)} \tau_{i_1}^z \tau_{i_2}^z + \ldots \,\, ,
\end{equation}\label{eq:HEff}
where the $\tau_{i}^z$ are the Q-LIOMs and the $n$-th order interaction constants $\lambda_{i_1,\ldots,i_n}^{(n)}$ are expected to fall off exponentially with distance. For example, $\lambda_{i_1,i_2}^{(2)} \sim e^{-d(i_1,i_2)/\xi}$, where $\xi$ is the localization length and $d(i_1,i_2)$ is the distance between site $i_1$ and site $i_2$. The predicted logarithmic growth of entanglement, the marker of a genuine MBL phase, is a direct consequence of the existence of the exponentially small non-local tails of the Q-LIOMs, and of their interaction~\cite{Serbyn2013,Ros2015}. Measuring the slow growth of entanglement entropy which is responsible for the unique structure of MBL phases is extremely challenging, although recent progress has been made~\cite{Lukin256}. This is mainly due to the highly non-local character of the half-chain entanglement entropy, which is not an easily measurable quantity beyond the small systems \cite{islam2015measuring,Lukin256}. Such difficulty inspired alternative ways to witness such dynamical behavior~\cite{Campbell2017,Iemini2016,Serbyn2014a,DeTomasi2017,Bera2016,Goold2015,Serbyn2014}.

The purpose of this letter is to demonstrate that the logarithmic spread is encoded in the behavior of the single-site density matrices and the local total correlations when an initial state for propagation is carefully chosen. We believe that this behaviour is experimentally accessible with the available techniques of local tomography; thus, we argue that local measurements are able to distinguish Anderson Localisation (AL) from its interacting counterpart i.e. MBL. This provides a strategy for experimental detection of unique MBL phenomenology. The letter is organized as follows. First, we provide a theoretical argument supporting a logarithmic growth of the local entropy and its relation to the notion of total correlations. Secondly, we discuss the optimal initial states, drawn from the HUB, to use in the time evolution. These special states allow us to obtain the longest transient dynamics in the localized state (for systems of finite size). Finally, we give numerical results for the local entropy and total correlations in the time evolution of the model \eqref{eq:XXZ}.

\paragraph{Logarithmic growth of entanglement.}
Let us consider the one-dimensional, disordered, isolated quantum system of $L$ spin-$1/2$ defined by Eq. \eqref{eq:XXZ}. Calling $\ket{\psi_t}$ the state of the whole system at time $t$, the reduced state of the $n$th site is obtained by tracing out the complement $L/n$: $\rhont \coloneqq \Tr_{L / n} \ket{\psi_t} \bra{\psi_t}$. The bipartite entanglement between the $n$th site and the rest is quantified by the Von Neumann entropy of the reduced state:
\begin{equation}
S_n(t) \coloneqq - \Tr \rhont \log \rhont 
\end{equation}
This quantity is experimentally accessible by means of local tomography. Moreover, one can also consider the average entropy over all sites:
\begin{equation}
S(t) = \frac{1}{L}\sum_{n=1}^L S_n(t) \, .
\label{eq:average_entropy}
\end{equation}

The latter quantity has the following operational meaning. Let $\mathcal{P} \subset \mathcal{H}$ be the set of all tensor product states of an $L$-partite quantum system. The total correlations $T(\rho)$ of a (possibly mixed) state $\rho$ is defined as~\cite{modi2010unified}
\begin{equation}
T(\rho) = \min_{\pi \in \mathcal{P}}S(\rho |\!| \pi) \, ,
\end{equation}
where $S(\rho |\!| \sigma)\coloneqq - \Tr \rho \log \sigma - S(\rho)$ is the relative entropy. 
$T(\rho)$ is an extensive quantity that measures the distinguishability between $\rho$ and the closest product state $\pi_\rho \in \mathcal{P}$. It turns out that, for each $\rho$, such state is unique. It is the product state of the reduced density matrices obtained from $\rho$: $\pi_\rho = \rho_1 \otimes \ldots \otimes \rho_L$ where $\rho_i \coloneqq \Tr_{L/i}\rho$. In our case of a pure state $\ket{\psi_t}$ it is easy to see that the total correlations $T_t\coloneqq T(\ket{\psi_t}\bra{\psi_t})$ are simply a rescaling of $S(t)$
\begin{equation}
T_t = \sum_{n=1}^L S_n(t) - S(\ket{\psi_t}\bra{\psi_t}) = L S(t) \label{eq:Tot2}
\end{equation}
It was recently shown that the study of the total correlations in the diagonal ensemble can signal the transition from ergodic to the MBL phase~\cite{Goold2015, Pietracaprina2017} but since the diagonal ensemble is a mixed state this requires knowledge of the global state. Since the states here are pure the total correlations can be probed dynamically by means of only local operations. 

In Ref.~\cite{Serbyn2013}, an argument has been proposed to explain the logarithmic growth of the bipartite entanglement entropy $S_{L/2}(t)$ that was previously numerically observed in Ref.~\cite{Bardarson2012}. Here we summarize this argument and explore its consequences for the growth of the local entropy. The intuition is based on the presence of the exponentially suppressed tails of the Q-LIOMs, which decay on a length scale given by the localization length $\xi$.  Calling $\Delta$ the coupling constant of the $s^z_i s^z_{i+1}$ interaction term, if there are no interactions ($\Delta=0$) all energy eigenstates are single-particle excitations, and there is Anderson Localization. In presence of interactions we have many-body localization. In this case, if two particles are placed at a distance $x_{ij}$ they would have an interaction energy which is exponentially suppressed because of the exponentially small tails $V_{ij} \sim \Delta e^{-x_{ij}/\xi}$. The dephasing time between them is therefore $t_{ij}\sim \hbar/V_{ij}=\hbar e^{x_{ij}/\xi}/\Delta$. This implies that $S_{L/2}(t)$ should grow in time with a logarithmic law, as outlined in Refs.~\cite{Serbyn2013, Bardarson2012}.

We now look at the implications of this argument for the bipartite entanglement between a single site and the rest and hence for the total correlations. The degrees of freedom on a lattice site $n$ will become entangled with the degrees of freedom living on the $n+k$ site on a time-scale $t_k \sim t_{\mathrm{min}}e^{k a/\xi}$ where $a$ is the lattice spacing. As time evolves the $n$-th site will become entangled with an increasing number of sites. The higher the number of sites which have entanglement with the $n$-th one, the higher $S_n(t)$, which will accumulate on a logarithmic time scale. From this we expect a logarithmic growth of $S_n(t)$. We would like to stress here that this argument holds for each individual site, thus requiring probing only a single site (especially important e.g. in an experimental setup). Although in the following we will show results for the average single site entanglement entropy \eqref{eq:average_entropy}, we checked that the result for each individual sites is quantitatively the same.
$S(t)$ quantifies the average growth of bipartite entanglement between one site and the rest of the chain. Since $S(t)$ is a rescaling of the total correlations, it additionally provides an upper bound to the amount of multipartite entanglement present in the system once rescaled with the system size.

\paragraph{Initial states.} In the study of the dynamics of isolated quantum systems, an crucial ingredient is the choice of the initial state. A typical criterion driving this choice is experimental feasibility; a well-known example is the anti-ferromagnetic (N\'eel) state, which can be prepared as the ground state of a local Hamiltonian. From the information-theory point of view the N\'eel state (polarized e.g. along the $z$ direction) is part of the so-called computational basis $\mathcal{B}_z \coloneqq \left\{ \ket{s^z_1} \ket{s^z_2}\ldots \ket{s^z_L}\right\}$, the tensor-product basis of the $z$ components of the local spins  $\ket{s^z_i} \in \left\{ \ket{\uparrow}_z,\ket{\downarrow}_z\right\}$. Here, we additionally request that the chosen initial state allows for sufficiently long dynamics in the MBL phase before reaching saturation. Since our goal is to probe dynamical features of an isolated quantum system, our initial state should not be too close to being a single energy eigenstate. Indeed, if this was the case, the state dynamics would always occur in the proximity of the initial state, resulting in rapid dynamics for observables before saturation that is unlikely to be seen. The choice of initial state is therefore crucial and by starting with a state which is a superposition of as many energy eigenstates as possible, we can access a longer transient dynamics which explores a larger part of the Hilbert space.

We focus on the strong disorder regime where the system is characterized by an extensive number of Q-LIOMs and the energy eigenstates are close to elements of the computational basis along the $z$ direction (basis in which the magnetic field term in Eq. \eqref{eq:XXZ} is diagonal). Therefore, in the MBL phase, the N\'eel state polarized along the $z$ direction is very close to an energy eigenstate and has very short dynamics before saturation.
A set of initial states which can be used to avoid this issue is given by the elements of a Hamiltonian Unbiased Basis~\cite{Anza2017,Anza2018} (HUB). A basis $\mathcal{B}\coloneqq \left\{ \ket{v_{\mu}} \right\}_{\mu=1}^D$ is called a HUB when
\begin{equation}
\left\vert \braket{v_{\mu}}{E_{\nu}}\right\vert^2 = \frac{1}{D} \qquad \forall \,\, \mu, \nu \,\,\,,
\end{equation}
where $\ket{E_{\nu}}$ are the Hamiltonian eigenstates and $D$ is the dimension of the Hilbert space. 
If our initial state is part of a HUB, its decomposition in the Hamiltonian basis will include all eigenstates and will be as far away as possible from being an Hamiltonian eigenstate. In the case of the XXZ model with disordered magnetic field along the $z$ direction, deep in the MBL phase the Q-LIOMs will be almost diagonal in $\mathcal{B}_z$. Hence, deep in the MBL phase, states that are a tensor product of the local spins polarized along the $x$ or $y$ direction are close to HUBs:
\begin{align}
& \mathcal{B}_x \coloneqq \left\{ \ket{s_1^x}\ldots \ket{s_L^x} \right\}  &&  \mathcal{B}_y \coloneqq \left\{ \ket{s_1^y}\ldots \ket{s_L^y} \right\} \, , \label{eq:HUBs}
\end{align}
where $\ket{s_i^x}\in \left\{ \ket{\uparrow}_x,  \ket{\downarrow}_x\right\}$ and $\ket{s_i^y}\in \left\{ \ket{\uparrow}_y,  \ket{\downarrow}_y\right\}$ for all $i=1,\ldots,L$. 

Here we will consider initial states that are elements of the bases $\mathcal{B}_x$ and $\mathcal{B}_y$, and specifically the N\'eel state along the $x$ direction $\ket{\uparrow\downarrow\uparrow\downarrow\ldots}_x$. These states have contributions from all subspaces that would conserve total $S_z$ magnetization and have already been prepared and used for experiments in Refs.\cite{Wei2018,Smith2016,Zhang2017,Liu2019,Hess2017,Zhang2017a}. The results do not depend on this  choice, and in the Supplementary materials we show that the N\'eel state along the $y$ direction $\ket{\uparrow\downarrow\uparrow\downarrow\ldots}_y$, the ferromagnetic states $\ket{\uparrow\uparrow\uparrow\uparrow\ldots}_x$ and $\ket{\uparrow\uparrow\uparrow\uparrow\ldots}_y$ and the states with two polarized domains $\ket{\uparrow\uparrow\ldots\downarrow\downarrow\ldots}_x$ and $\ket{\uparrow\uparrow\ldots\downarrow\downarrow\ldots}_y$ along the $x$ and $y$ directions all give the same logarithmic growth and phenomenology outlined in the next paragraph.

\begin{figure}[ht]
\includegraphics[width=\linewidth,height=.7\linewidth]{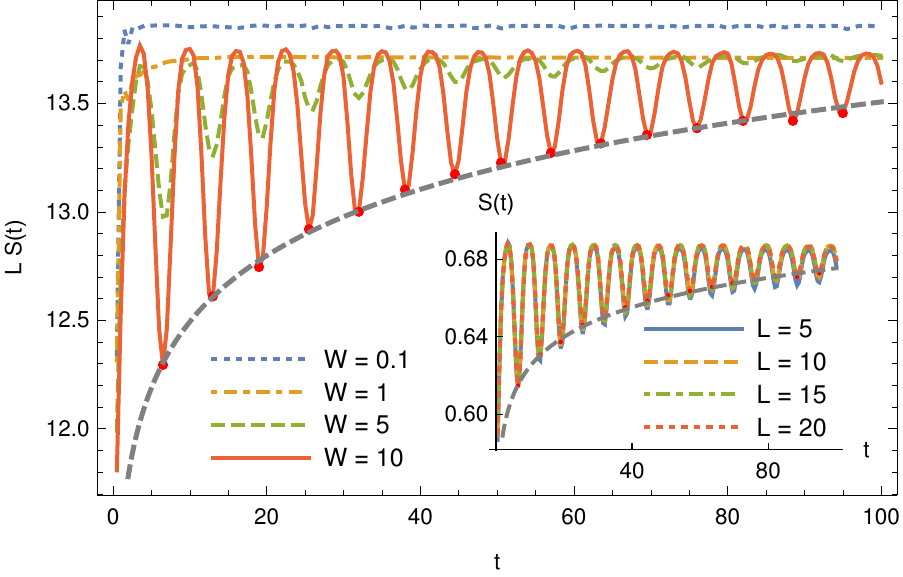}
\caption{Time-dependent behaviour of $S(t)$ for $L=20$ and $W=0.1,1,5,10$. In the MBL phase $S(t)$ is modulated on a logarithmic time-scale (the dashed line is a logarithmic fit of the minima). The inset shows the behaviour of $S(t)$ at $W=10$ for $L=20$, compared to the one for system sizes $L=5,10,15$, showing very weak system size dependence up to very small sizes. Further checks have been performed with systems of size $L=5$ through $L=20$, all showing quantitatively the same time-dependent profile.}\label{Figure1}
\end{figure}

\paragraph{Results.} We consider the time evolution of the spin chain \eqref{eq:XXZ} with $\Delta=1$ and four values of the disorder strength in the delocalized (very low $W=0.1$ and low $W=1$ disorder) and localized ($W=5$ and $W=10$) phases. We consider systems up to $L=20$ spins, averaging over a sufficient number of disorder realizations ($1000$, $100$ and $50$ realizations for sizes $L\leq12$, $13\leq L\leq19$, $L=20$ respectively). To perform the time evolution, we used an iterative Krylov subspace method with the Lanczos algorithm that avoids full diagonalization~\cite{sidje1998expokit, Varma2017a}.

\begin{figure}[bht]
\includegraphics[width=\linewidth]{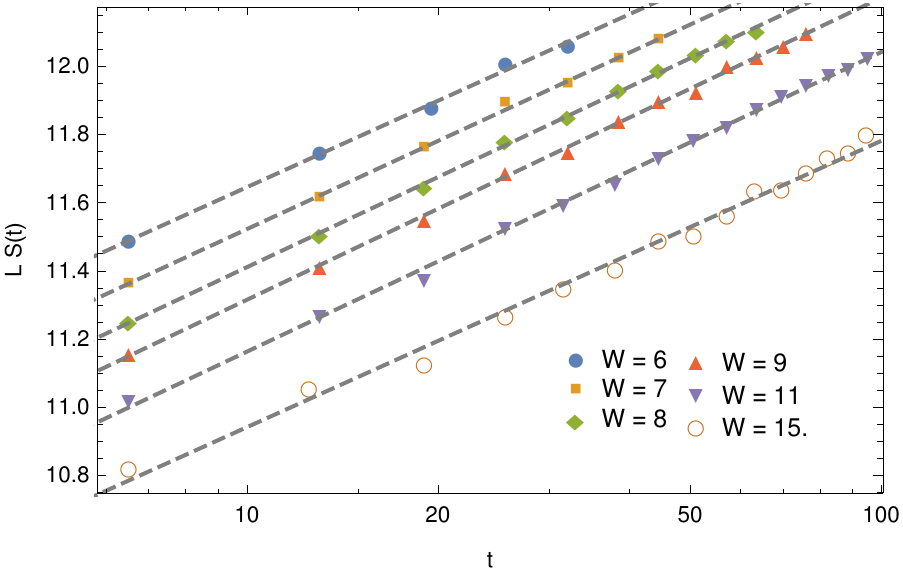}
\includegraphics[width=\linewidth]{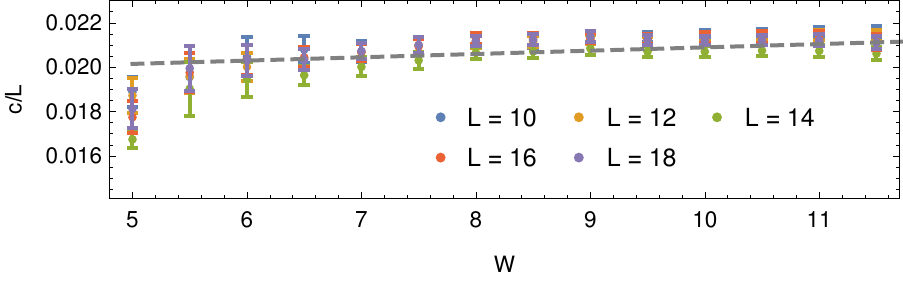}
\caption{Logarithmic growth of $S(t)$ for different disorder strengths. Top panel: Local minima (obtained as in Fig. \ref{Figure1}) as a function of time for size $L=18$. Bottom panel: Coefficient $c$ of the fit $L S(t)=a+c \, \log (t)$ as a function of disorder strength, rescaled by system size $L$, for values $L=10,12,14,16,18$.}\label{Figure2}
\end{figure}

With the initial states outlined above, we clearly obtain a logarithmic growth for the single-site entanglement entropy: in Fig. \ref{Figure1} we plot the behaviour of $S(t)$ at $L=20$ for the different values of $W$, showing that a logarithmic envelope  is present only in the MBL phase ($W=5,10$). In the thermal phase ($W=0.1,1$) we simply observe a quick thermalization towards the maximum entropy configuration. The inset shows that the same behavior emerges already for sizes as small as $L=5$. More details can be found in Section \ref{sm:Peaks} of the Supplemental materials.
At each $L$ we can isolate the local minima of $S_{i}(t)$ and extract the slope $c(W)$ of the logarithmic growth as function of $W$. The data are shown in Fig.\ref{Figure2}, showing that $c/L$ is constant within the error. Understanding the relation of $c$ with quantities of phenomenological relevance, as the localization length, goes beyond the purpose of this letter and it is left for future investigation.
\begin{figure}[bht]
\includegraphics[width=\linewidth,height=.6\linewidth]{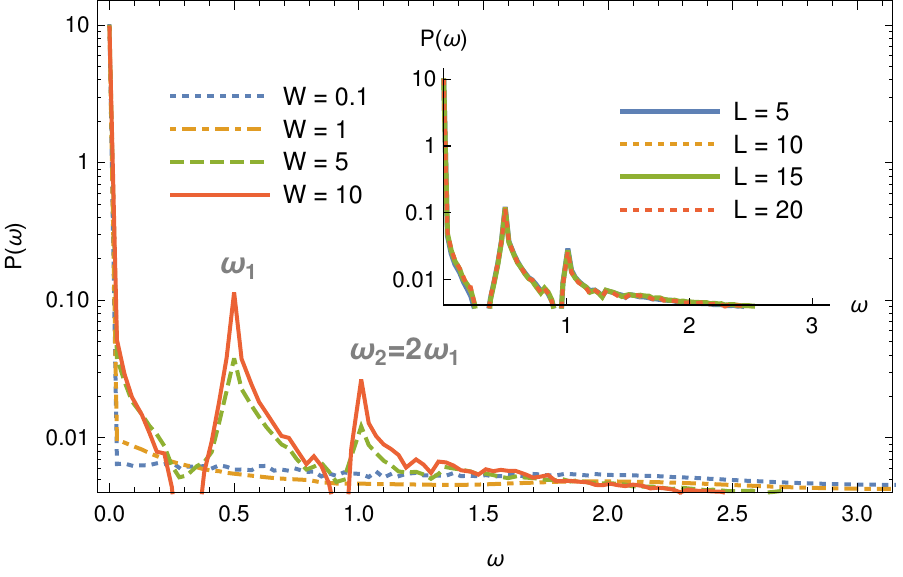}
\caption{Power spectrum of $S(t)$ for $L=20$ and $W=0.1,1,5,10$. There are two visible peaks, whose frequencies are related by $\omega_2=2\omega_1$. The inset shows the power spectrum of $S(t)$, at $W=10$, for $L=5,10,15,20$, which is quantitatively independent on the size of the system.}\label{Figure2Bis}
\end{figure}

Due to its quasi-periodic behaviour, we performed a discrete Fourier transform $\mathcal{F}(\omega)$ on $S(t)$  and studied its power spectrum $P(\omega):=|\mathcal{F}(\omega)|$  to understand its oscillations.
In Fig. \ref{Figure2Bis}, we show the behavior of the power spectrum for $W=10$ and different sizes $L=5,\ldots,20$. As for $S(t)$, its structure is  independent on the size of the system. The positions of the first ($\omega_1$) and the second ($\omega_2$) peaks are related to each other by a simple relation $\omega_2=2 \omega_1$. This is due to the fact that, in the MBL phase, the eigenvalues $\lambda_{\pm}^{(n)}(t)$ of $\rho_t^{(n)}$ have a periodic structure which is modulated on a logarithmic time-scale. Because of that, when we perform the Fourier transform of $S_n(t)$, the terms in the Taylor expansion of $S_n(t)$ as a function of $\lambda_{\pm}^{(n)}(t)$ are responsible for the presence of peaks at frequencies that are multiple integers of the lowest frequency: $\omega_n = n \omega_1$. In Figure \ref{Figure2Bis} only the first two are visible.

Finally, we remark once again that all the $S_n(t)$ have the same time-dependent profile and power spectrum as their average $S(t)$. 

\begin{figure}[htbp]
\includegraphics[width=1\linewidth]{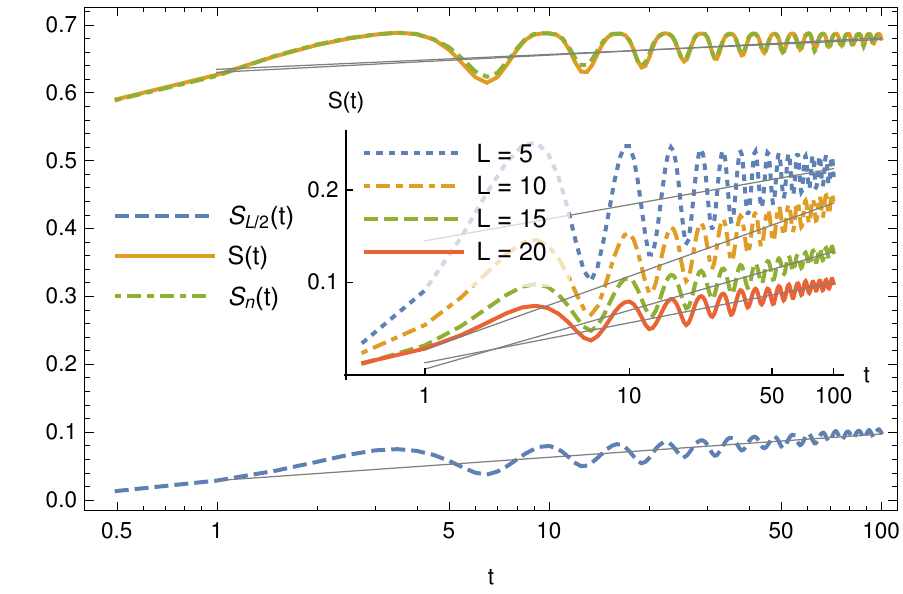}
\caption{Time-dependent behaviour of the half chain entropy $S_{L/2}(t)$ compared to $S(t)$ and the local $S_n(t)$, for $L=20$ and $W=10$. The inset shows how $S_{L/2}(t)$ behaves for $L=20,12,6$ and $W=10$. We found no system size dependence: sizes $L=5$ through $20$ exhibit the same time-dependent profile.}\label{fig:HalfChain}
\end{figure}

As a final check of the consistency of the local entropy results with the known $\log(t)$ behavior of the half-chain entropy $S_{L/2}(t)$, in Fig. \ref{fig:HalfChain} we compare their behaviour for strong disorder $W=10$, showing that they both exhibit the logarithmic growth in time.

\begin{figure}[htbp]
\includegraphics[width=\linewidth]{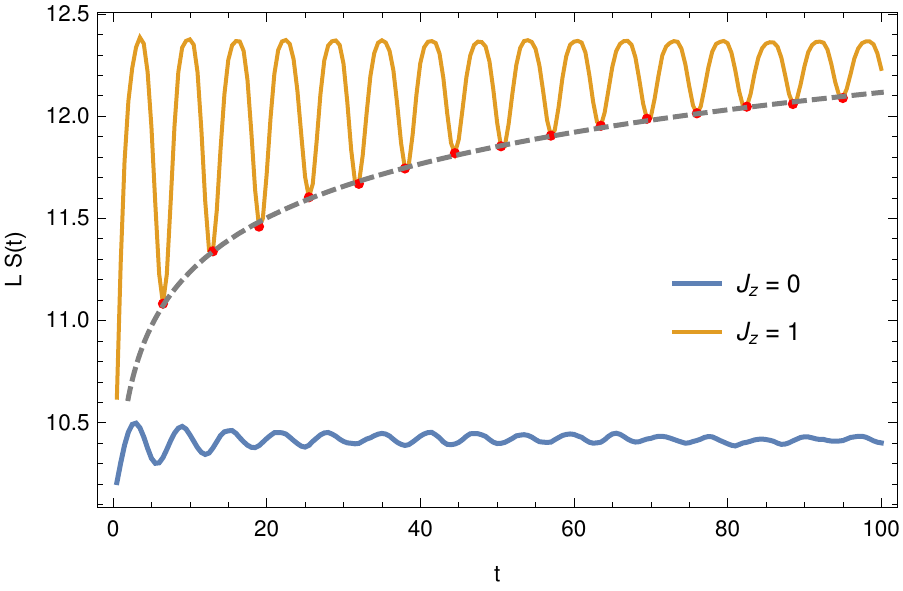}
\caption{Time-dependent behaviour of $S(t)$ for system size $L=18$ and disorder $W=10$ for the interacting ($J_z=1$) and the non-interacting ($J_z=0$) systems, showing the different behavior of the MBL phase (logarithmic growth, Q-LIOMs oscillations) and the AL phase (saturation, damped oscillations).}\label{fig:noninteracting}
\end{figure}

\paragraph{Discussion.} We stress that the logarithmic spread of entanglement is one of the characteristic features of genuine MBL, as opposed to AL: indeed, in Fig. \ref{fig:noninteracting} we show the two qualitatively different behaviors in the time evolution of our model \eqref{eq:XXZ}. Experimentally, the detection of genuine MBL through the measurement of entanglement is a daunting proposition. Here we have demonstrated that this definitive signature of MBL can be obtained from local measurements alone. We obtained this result through a careful choice of the initial state, building on the notion of Hamiltonian Unbiased Basis (HUB), and through both a theoretical argument and numerical evidence that the logarithmic spread of entanglement is encoded in the behavior of each individual single-site entropies.

There are several unique features of the analysis that deserved to be highlighted. Firstly, the logarithmic modulation in time of the single-site entanglement entropy is quite evident for systems of sizes as small as $L=5$, very far away from the thermodynamic limit. Secondly, the study of the power spectrum of $S_n(t)$ (and $S(t)$) shows that the presence of the  peaks in Fig.\ref{Figure2} is a distinctive feature of the MBL phase. Again, this is clear already for systems of size as small as $L=5$. This points towards the fact that the power spectrum of the entanglement entropy can be a useful tool to investigate quantum systems, beyond its relevance for the physics of MBL systems.

{Finally, we highlight once more that a number of experimental setups are especially suited to measure the local quantities that we propose here; these are trapped ions~\cite{exp_4} and nuclear magnetic resonance setups~\cite{exp_5}.}

\paragraph{Acknowledgements.} F.A. would like to acknowledge  many discussions on the phyiscs of MBL systems with X. Lei and J. Crutchfield. F.A. acknowledges that this project was made possible through the support of a grant from Templeton World Charity Foundation, Inc.. The opinion expressed in this publication are those of the authors and do not necessarily reflect the views of Templeton World Charity Foundation, Inc.
F.P. acknowledges the support of the project THERMOLOC ANR-16-CE30-0023-02 of the French National Research Agency (ANR) and thanks F. Alet for useful suggestions. This work was supported by an SFI-Royal Society University Research Fellowship (J.G.). This project received funding from the European Research Council (ERC) under the European Union's Horizon 2020 research and innovation program (grant agreement No.~758403.

\bibliographystyle{plainnat}
\bibliography{library}

\onecolumn\newpage
\appendix

\section{Initial states, Hamiltonian Unbiased Basis and Power Spectrum}\label{sm:InitialStates}

As stated in the main text, we prepare the system in an initial state which is close to an element of an Hamiltonian Unbiased Basis (HUB). 
This offers a substantial advantage in our case: the initial state will be a linear superposition of as many energy eigenstates as possible, allowing both a longer and a wider exploration of the Hilbert space.

\begin{figure}[ht]
\begin{center}
\includegraphics[width=0.5\linewidth]{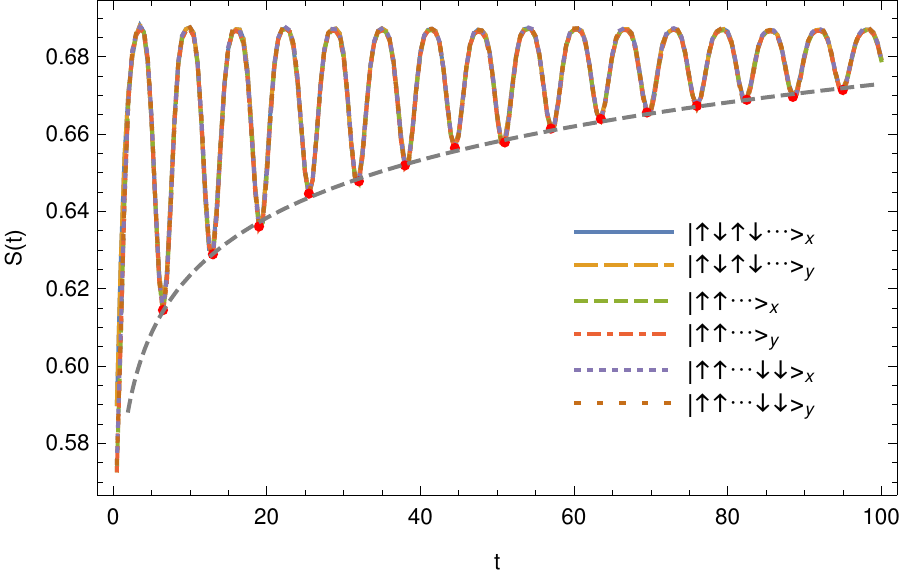}
\caption{Average local entropy $S(t)$ as a function of time for various initial states that closely approximate a member of a HUB at strong disorder, for a system of size $L=18$ and disorder $W=10$.
}\label{fig:hub}
\end{center}
\end{figure}

In the left panel of Fig.~\ref{fig:hub} we show that the N\'eel state along the $x$ direction closely approximates a HUB at high disorder, in contrast with the N\'eel state along the $z$ direction which has overlap only with very few eigenstates.

In the main text we only showed the results obtained with the initial state $\ket{\uparrow\downarrow\uparrow\downarrow\ldots}_x$.
In the right panel of Fig. \ref{fig:hub} we show that the results are quantitatively the same if we consider any one of the following additional initial states, all of which closely approximate a HUB: $\ket{\uparrow\downarrow\uparrow\downarrow\ldots}_y$, $\ket{\uparrow\uparrow\uparrow\uparrow\ldots}_x$, $\ket{\uparrow\uparrow\uparrow\uparrow\ldots}_y$, $\ket{\uparrow\uparrow\ldots\downarrow\downarrow\ldots}_x$ and $\ket{\uparrow\uparrow\ldots\downarrow\downarrow\ldots}_y$.

With all the initial states outlined above, we clearly obtain a logarithmic growth for the single-site entanglement entropy. Moreover, in Fig. \ref{Figure1} of the main paper we plotted the behaviour of $S(t)$ at $L=20$ for different values of $W$, showing that a logarithmic envelope (red line) is present only in the MBL phase ($W=5,10$).

\section{Lattice-variance of the single-site entanglement entropy}\label{sm:Variance}

\begin{figure}[htb]
\begin{center}
\includegraphics[width=.5\linewidth]{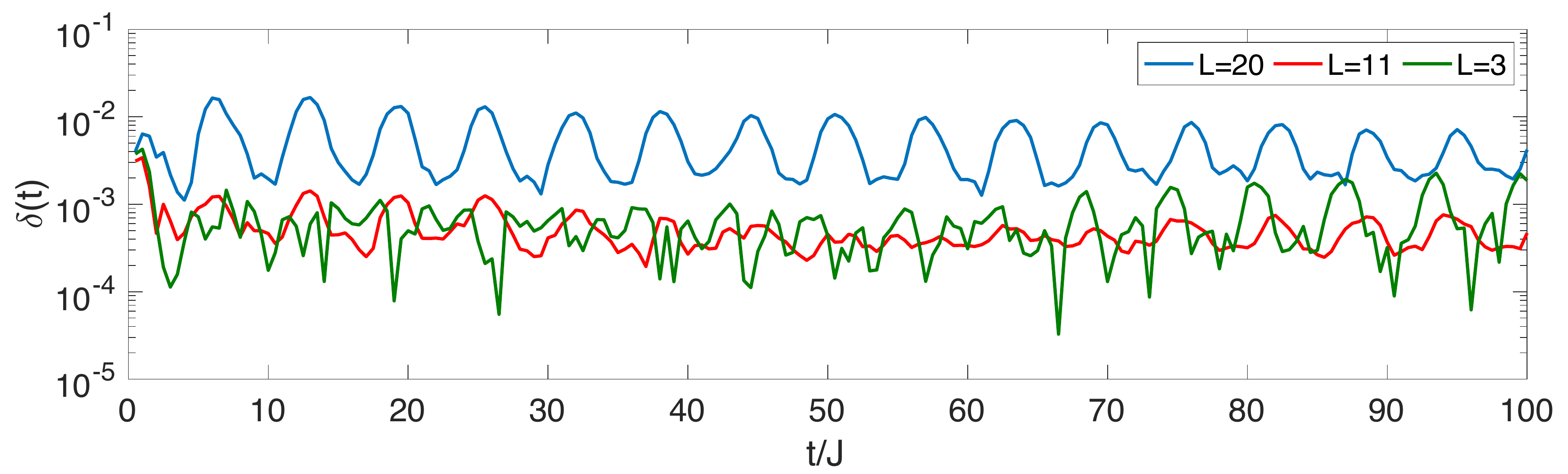}
\caption{Variance of the local entropy across the lattice (Eq.\ref{eq:variance}), for three system sizes $L=20,11,3$ at $W=10$.}\label{fig:Variance}
\end{center}
\end{figure}

In the main text we claim that the local entropies $S_n(t)$ and their average $S(t)$ are quantitatively the same behavior. To show this, in Figure \ref{fig:Variance} we plot the  square root of the average difference-squared between the local entropies $S_n(t)$ and $S(t)$
\begin{equation}
\delta(t) = \sqrt{\frac{1}{L} \sum_{n=1}^L (S_n(t)-S(t))^2} \,\,\,, \label{eq:variance}
\end{equation}
This quantity is the lattice-variance of the single-site entanglement entropy. It estimates, at each time, how much each of the $S_n(t)$ differ from their average $S(t)$. As we can see in Figure \ref{fig:Variance}, this appears to be consistently small for all the parameters of the model and at any time. Indeed, we find that $\delta(t) \lesssim 10^{-3}$, for all system sizes, values of the disorder and at any time.

\section{Position of the peaks in the Fourier transform of the single-site entropy}\label{sm:Peaks}

In Figure \ref{Figure2Bis} of the main text we plot the power spectrum of the entanglement entropy, which has a peculiar behavior in the MBL phase.

We observe the emergence of two major peaks at frequencies, $\omega_1$, $\omega_2$ which are related via $\omega_2 = 2 \omega_1$. Here we provide an analytic argument to understand why there is such relation.

The first step is to notice that the behavior emerges already for very small system sizes. This, together with the fact that the period of the oscillation is relatively small, suggests that the existence of  the oscillations should be due to short-range interactions between nearest neighbour spins. To verify this idea we choose $L=10$ and $W=10$ and studied what happens to the power spectrum of the average single-site entropy $S(t)$ when $\Delta$ changes. In particular we focus on $\Delta=0.2,0.4,0.6,0.8,1$.

\begin{figure}[htb]
\begin{center}
\includegraphics[width=0.9\linewidth]{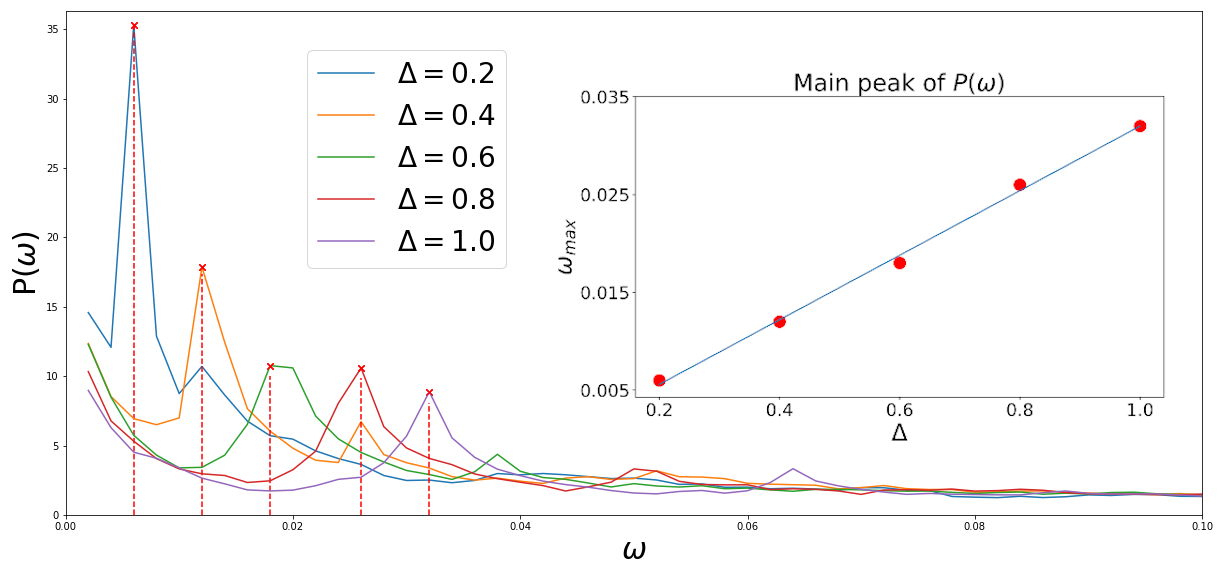}
\caption{Here we show how the power spectrum of the average single-site entropy, $S(t)$, changes as a function of the interaction strength $\Delta$, for a system with $L=10$ spins and $W=10$. In the main figure we show the various power spectra at different values $\Delta = 0.2, 0.4, 0.6, 0.8, 1$. In the inset we plot the location of the largest peak of the power spectrum as a function of $\Delta$, which exhibits a linear dependence.  }\label{fig:Oscillations}
\end{center}
\end{figure}

As we can see from Figure \ref{fig:Oscillations} the frequency of the oscillations is determined by the interaction strength $\Delta$, which in the main text is fixed to $1$. Intuitively, the existence of the localized phase suppresses the hopping term favouring the magnetic field term and the $\sigma_z-\sigma_z$ nearest neighbour interaction, whose intensity is regulated by $\Delta$. For this reason, we believe the behaviour of the oscillations can be understood by looking at the following two-spins model with phenomenological Hamiltonian 
\begin{equation}
H_{\mathrm{phen}} = \sigma_1^z + \sigma_2^z + V_{\mathrm{int}} \,\, \sigma_1^z \sigma_2^z 
\end{equation}

\begin{figure}[htbp]
\includegraphics[width=0.9\linewidth]{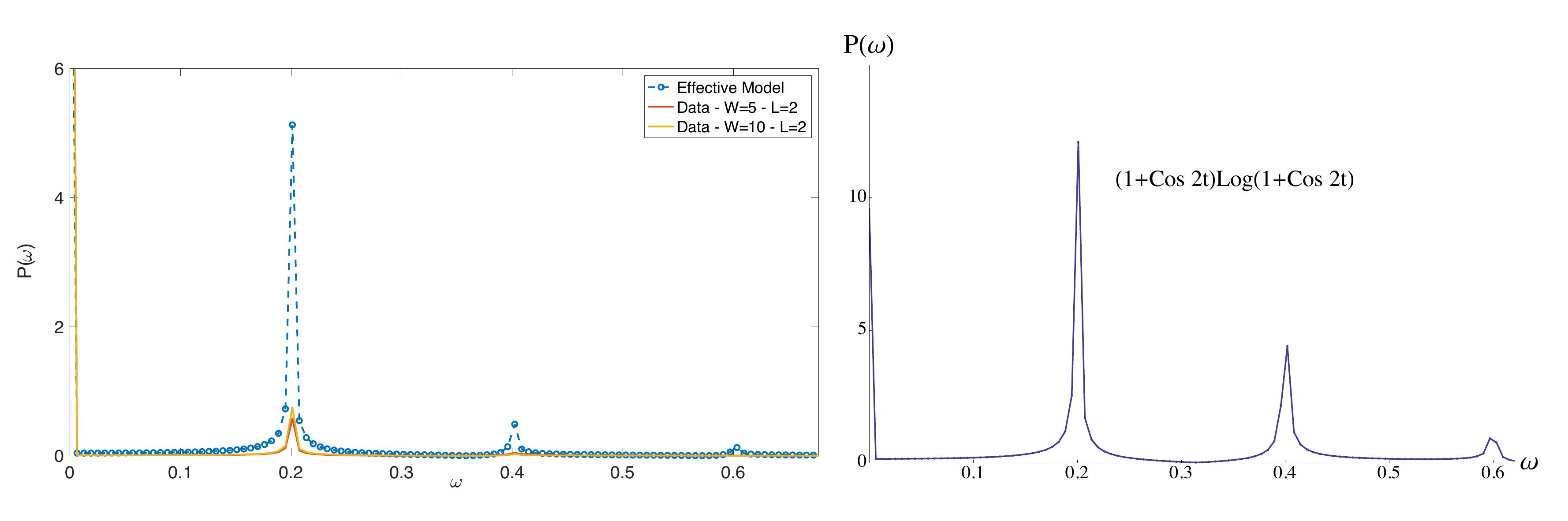}
\caption{Left panel: Comparison between the power spectrum of the entanglement entropy that we obtain from the effective model and the one that we obtain from the data on the Heisenberg model at $L=2$. All the peaks are located at the same position for $V_{\mathrm{int}}=1/4$. The precise value of $V_{\mathrm{int}}$ is empirical. 
Right panel: Power spectrum of the function $f(t)=(1+\cos 2t) \log (1+\cos 2t)$. The secondary peaks are located at frequencies which are multiple integers of the frequency of the major peak.}\label{fig:Peaks}\label{fig:Peak_Example}
\end{figure}

The dynamics of such simple theoretical model can be solved exactly. In particular, the Hamiltonian has three distinct eigenvalues: $E_{\mathrm{min}}$, $E_{\mathrm{max}}$ and $E_{\mathrm{C}}$. The central one, $E_{\mathrm{C}}$, has degeneracy $2$. So we have

\begin{align}
& E_{\mathrm{max}} = 2 + V_{\mathrm{int}} && E_{\mathrm{C}} = -V_{\mathrm{int}} &&& E_{\mathrm{min}} = -2 + V_{\mathrm{int}}
\end{align}
Respectively, their eigenstates are
\begin{align}
&\ket{E_{\mathrm{max}}} = \ket{\uparrow \uparrow} &&\ket{E_{\mathrm{C}}(\alpha,\theta)} = \alpha \ket{\uparrow \downarrow} + \sqrt{1-|\alpha|^2}e^{i\theta} \ket{\downarrow \uparrow} &&&\ket{E_{\mathrm{min}}} = \ket{\downarrow \downarrow}
\end{align}
We can easily write down the propagator:
\begin{equation}
U(t) = e^{- \frac{i}{\hbar}H_{\mathrm{eff}}t}= e^{-\frac{i}{\hbar}(2+V_{\mathrm{int}})t} \ket{\uparrow \uparrow}\bra{\uparrow \uparrow} + e^{\frac{i}{\hbar}V_{\mathrm{int}}t} \left( \ket{\uparrow \downarrow}\bra{\uparrow \downarrow} + \ket{\downarrow \uparrow}\bra{\downarrow \uparrow} \right) + e^{-\frac{i}{\hbar}(-2+V_{\mathrm{int}})t} \ket{\downarrow \downarrow} \bra{\downarrow \downarrow}
\end{equation}
As in our simulation, the initial state that we choose is the Neel state, polarized along the $X$ direction:
\begin{equation}
\ket{\psi_0} = \ket{\mathrm{Neel}_X} = \ket{\downarrow_x,\uparrow_x} = \frac{1}{\sqrt{4}} \left( \ket{\uparrow \uparrow}+\ket{\uparrow \downarrow}-\ket{\downarrow \uparrow}-\ket{\downarrow \downarrow}\right)
\end{equation}
Which gives the following exact expression for the time-dependent pure state:
\begin{equation}
\ket{\psi_t} = \frac{1}{\sqrt{4}} \left[ e^{-\frac{i}{\hbar}(2+V_{\mathrm{int}})t}\ket{\uparrow \uparrow}+ e^{-\frac{i}{\hbar}V_{\mathrm{int}} t}\left(\ket{\uparrow \downarrow}-\ket{\downarrow \uparrow} \right)-e^{-\frac{i}{\hbar}(-2+V_{\mathrm{int}})t}\ket{\downarrow \downarrow}\right]
\end{equation}
Eventually, the time-dependent density matrix is obtained via the outer product of $\ket{\psi_t}$ with $\bra{\psi_t}$:

\begin{equation}
\rho_t =
\frac{1}{4}\left( \begin{array}{cccc}
1 & e^{-\frac{i}{\hbar}(2+2V_{\mathrm{int}})t} & -e^{-\frac{i}{\hbar}(2+2V_{\mathrm{int}})t} & -e^{-\frac{i}{\hbar}4t} \\
 c.c. & 1 & -1 & -e^{-\frac{i}{\hbar}(2-2V_{\mathrm{int}})t} \\
 c.c & c.c. & 1 & -e^{-\frac{i}{\hbar}(2-2V_{\mathrm{int}})t} \\
 c.c. & c.c. & c.c. & 1
\end{array} \right)
\end{equation}
From this, it is easy to compute the partial trace over the firs spin and obtain the reduced density matrix of the second one. For the sake of simplicity, here we do not make explicit the value of the energy eigenvalues:
\begin{equation}
\rho_t^{(2)} =
\frac{1}{4}\left( \begin{array}{cc}
1+1 & e^{-\frac{i}{\hbar}(E_{\mathrm{max}}-E_C)t}+ e^{-\frac{i}{\hbar}(E_C - E_{\mathrm{min}})t}\\
e^{\frac{i}{\hbar}(E_{\mathrm{max}}-E_C)t}+ e^{\frac{i}{\hbar}(E_C - E_{\mathrm{min}})t} & 1+1
\end{array} \right)
\end{equation}
The Von Neumann entropy of this reduced state is the entanglement entropy. Hence, we need its eigenvalues $\lambda_2^{\pm}(t)$. After some algebraic manipulation and using the fact that $\Tr H_{\mathrm{eff}}=0$ we have
\begin{equation}
\lambda_2^{\pm}(t) = \frac{1 \pm \cos \theta(t)}{2}  \qquad \theta(t):=\frac{2|E_C|}{\hbar}t 
\end{equation}
Eventually, here is the analytical expression for the Entanglement entropy of a $L=2$ effective Hamiltonian model:
\begin{equation}
S_2(t) = - \frac{1+\cos \theta(t)}{2} \log \left( \frac{1+\cos \theta(t)}{2}\right) - \frac{1-\cos \theta(t)}{2} \log \left( \frac{1-\cos \theta(t)}{2}\right) \label{eq:vnentr}
\end{equation}
This is a function with smallest period $T = \frac{\hbar \pi}{2|E_C|}=\frac{\hbar \pi}{2|V_{\mathrm{int}}|}$. However, the entropy is a more complicated function, which involves the natural logarithm of $1+\cos \theta(t)$. The natural logarithm is responsible for the presence of peaks which go beyond the first one. Indeed, if we plot the Power spectrum of Eq.(\ref{eq:vnentr}) (see Figure \ref{fig:Peaks} right panel) with the ones obtained from the data, we see that the position of the peaks are in perfect agreement, for the choice $V_{\mathrm{int}}=1/4$. Incidentally, we notice that with such choice, $T = \frac{\hbar \pi}{2 V_{\mathrm{int}}} \approx 2\pi \hbar$.
Now, the reason why the secondary peaks are located at frequencies which are multiple integers of the frequency of the first peak seems to be purely technical. The logarithm is not a periodic function. However, the entropy in Eq.(\ref{eq:vnentr}) involves the logarithm of a periodic function. Thus, if we Taylor-expand the logarithm we obtain
\begin{equation}
\log (1+ \cos \theta(t)) = \sum_{n=1}^{\infty}(-1)^{n+1} \frac{[\cos \theta(t)]^n}{n} = \cos \theta(t) - \frac{[\cos \theta(t)]^2}{2} + \frac{[\cos \theta(t)]^3}{3}+ \ldots \label{eq:Taylor}
\end{equation}
Each one of these terms will give a peak which is centered in frequencies that are multiple integers of the original frequency of the $\cos \theta(t)$, due to the fact that we have integer powers of $\cos \theta(t)$. Indeed, if we look at the power spectrum of $(1+\cos 2t) \log (1+\cos 2t)$ we obtain the right panel of Figure \ref{fig:Peak_Example}.

\begin{figure}[htb]
\includegraphics[scale=0.6]{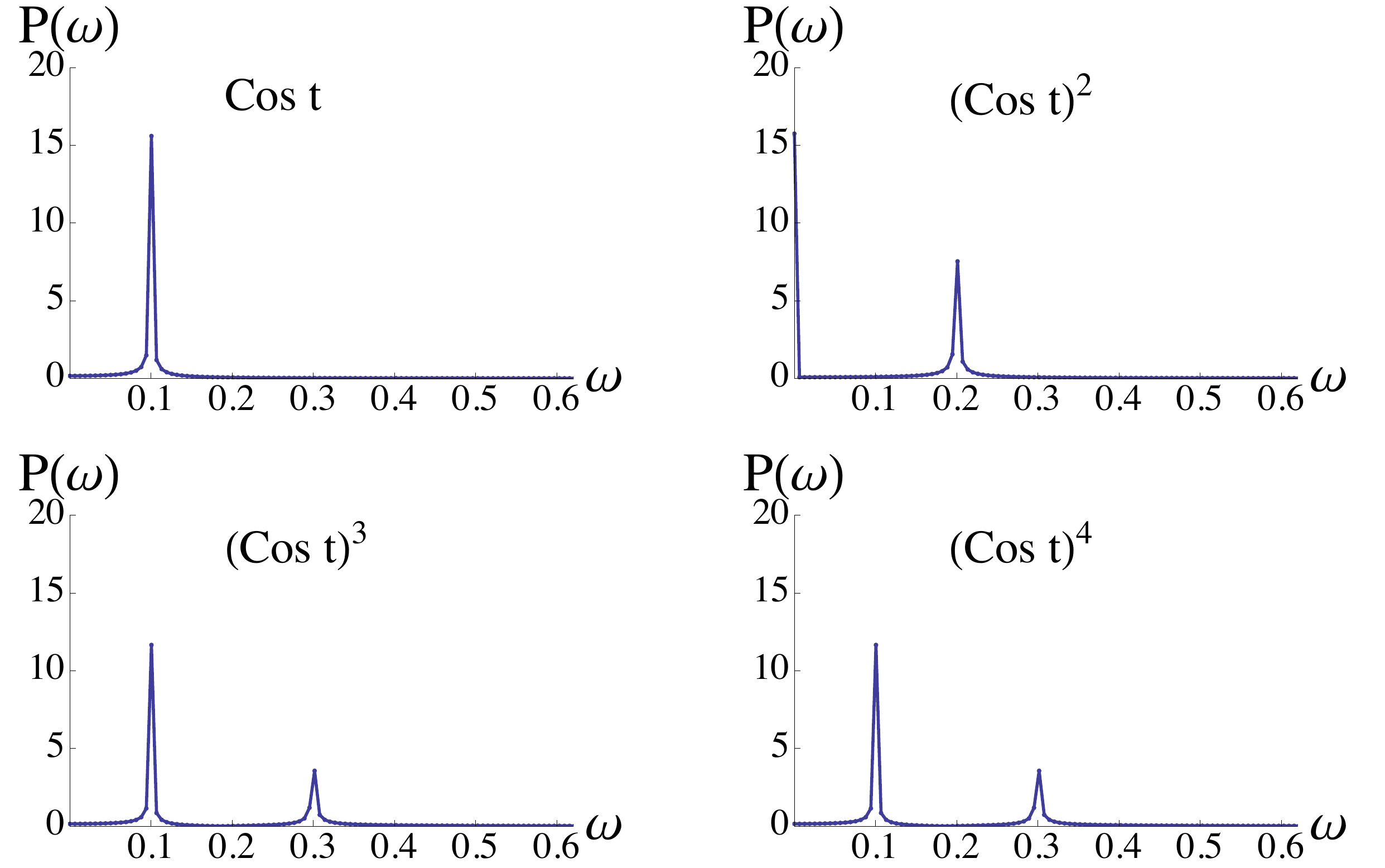}
\caption{Power spectrum of the functions $f_n(t)=(\cos t)^n$ for $n=1,\ldots,4$. The power spectrum of the entanglement entropy can be understood as a superposition of these objects, with certain coefficients.}\label{fig:Visualize_Peaks}
\end{figure}

More in general, if we forget for a moment about the coefficients, in the Taylor expansion of Eq.\ref{eq:Taylor}, using Euler formulas for the $\cos \theta$ we have
\begin{equation}
(\cos \theta)^n = \left( \frac{e^{i\theta}+e^{-i\theta}}{2}\right)^n = \frac{1}{2^n} \sum_{k=0}^n \binom{n}{k} e^{i\theta(n-k)} e^{-i\theta k} = \sum_{k=0}^n \frac{1}{2^n}\binom{n}{k} e^{i\theta (n-2k)}
\end{equation}
Hence, its Fourier transform is simply
\begin{equation}
\mathcal{F}[(\cos \theta)^n](\omega) = \sum_{k=0}^n \frac{1}{2^n}\binom{n}{k} \sqrt{2\pi} \delta(\omega-(n-2k))
\end{equation}
If we now consider that, in our case, $\theta$ is not the independent variable but it is linearly proportional to it, $\theta(t) = \frac{2|E_C|}{\hbar}t$, we conclude that the peaks will all be located at frequencies $\omega_n = \frac{2|E_C|}{\hbar}, 2\frac{2|E_C|}{\hbar}, 3\frac{2|E_C|}{\hbar}, \ldots, n\frac{2|E_C|}{\hbar}$
To help visualize the effect, in Figure \ref{fig:Visualize_Peaks} we plot the power spectrum of the first four powers of $\cos t$.

\end{document}